\documentclass[letter]{jpsj2}
\usepackage{amsmath}
\usepackage{amssymb}

\title{
Formation of non-unitary state near the upper-critical field of Sr$_2$RuO$_4$}

\author{
Masafumi \textsc{Udagawa}, Youichi \textsc{Yanase}, and Masao \textsc{Ogata}
}

\inst{Department of Physics, University of Tokyo, Hongo, Tokyo 113-0033, Japan}

\recdate{\today}

\abst{
We have studied the superconducting state of Sr$_2$RuO$_4$ under a
magnetic field parallel to the superconducting plane. We show that due
to a weak spin-orbit coupling, non-unitary
k$_y(\hat{\mathbf{z}}-i\alpha\hat{\mathbf{y}})$ state is formed
right at H$_{c2}$, then changes to unitary k$_y\hat{\mathbf{z}}$
state, as a magnetic field is lowered. In terms of this
crossover, we address the origin of the observed double
peaks of specific heat and the disappearance of the double peaks at low fields.
}

\kword{Sr$_2$RuO$_4$, two-component order parameter, specific heat,
tricritical point, non-unitary state
}

\begin{document}
\maketitle

Sr$_2$RuO$_4$ has been intensively studied in the last years, and its
superconducting properties are well understood as to a zero magnetic
field state\cite{rf:Maeno94,rf:Mackenzie03}. $\mu$SR experiment\cite{rf:Luke98} and a
microscopic calculation\cite{rf:Nomura02,rf:Yanase03} consistently support the chiral triplet
superconductivity with the order parameter, denoted as 
$\mathbf{d}(\mathbf{k})\propto(k_x\pm ik_y)\mathbf{\hat{z}}$.
Recently, momentum dependence of the gap amplitude was also determined with a
high-precision measurement of the angle-resolved specific heat\cite{rf:Deguchi04}.

However, the two-component order parameter
$\mathbf{d}(\mathbf{k})\propto(k_x\pm ik_y)\mathbf{\hat{z}}$ seems to
contradict a behavior under the
in-plane magnetic field. As Agterberg pointed out, the in-plane magnetic
field lifts the degeneracy between $k_x\mathbf{\hat{z}}$ and
$k_y\mathbf{\hat{z}}$\cite{rf:Agterberg98}. Given a magnetic field is applied parallel to
the {\it x}-axis, only one component with a shorter coherence length, say
$k_y\mathbf{\hat{z}}$, is condensed near the upper-critical field
($H_{c2}$). Since $k_y\mathbf{\hat{z}}$ is symmetric with respect to a
reflection about the {\it y}-axis while $(k_x\pm ik_y)\mathbf{\hat{z}}$
are not, a phase transition line must exist between {\it H} $=0$ and
$H=H_{c2}$ for all temperatures below $T_c$.

Actually, the observed double peaks of the specific
heat\cite{rf:Nishizaki00,rf:Deguchi02}, the thermal conductivity\cite{rf:Tanatar01}, and
the ac-susceptibility\cite{rf:Mao00} suggest an existence of a
second-order phase transition. However, these observations show several
inconsistencies with the prediction by
Agterberg\cite{rf:Agterberg98}. Firstly, the second peak was observed at much
higher field than the theoretical estimate. Secondly, the second peak line
terminates at ({\it H}, {\it T})$\sim$(1.2T, 0.8K) and merges into the $H_{c2}$
line. In particular, the termination of the second peak line suggests that
the zero-field $(k_x\pm ik_y)\mathbf{\hat{z}}$ state can be smoothly
continued up to H$_{c2}$ near T$_c$, thus inconsistent with the symmetry argument above.

Since the position of transition line is susceptible to the form of
the pairing function, etc., it may be possible to attribute the observed
peaks to Agterberg's phase transition\cite{rf:Kaur05}. However, in this paper, we would like
to address this problem from a different standpoint, i.e., by considering
the {\bf d}-vector degrees of freedom. In Sr$_2$RuO$_4$, at zero field, {\bf d}-vector
is fixed to be parallel to $\hat{\mathbf{z}}$ due to the spin-orbit
interaction between $3d$ electrons forming the three Fermi
surfaces\cite{rf:Yanase03}. However, the energy splittings between the stable (A)$(k_x\pm
ik_y)\mathbf{\hat{z}}$ and unstable (B)$k_x\mathbf{\hat{x}}\pm k_y\mathbf{\hat{y}}$ or 
(C)$k_x\mathbf{\hat{y}}\pm k_y\mathbf{\hat{x}}$ configurations are considerably
small. According to the microscopic calculation, it was estimated to be
less than 0.01$T_c$\cite{rf:Yanase03}. Such a small energy difference naturally
results from the fact that spin-orbit interaction affects the
superconductivity only in the order of
($\frac{\lambda}{W}$)$^2$, where $\lambda$ is the LS coupling constant,
and {\it W} is the band-width. These tiny energy differences are also
supported by the NMR experiment with the magnetic field parallel to
$\hat{\mathbf{z}}$\cite{rf:Murakawa04}. There the {\bf d}-vector is shown
to flip at $H \sim 200$Oe, which means the states (A)-(C) are degenerate within $\sim$0.001$T_c$.

First, let us explain qualitatively the physical mechanism of the double
peaks which is proposed in this paper. The quasi-degeneracy of order parameters enables the
$k_y\mathbf{\hat{y}}$ component to mix with the most stable
$k_y\mathbf{\hat{z}}$. Actually, under a magnetic field,
the mixing is caused by the spin-dependent chemical potential shift due
to Zeeman coupling with the external field. If the density of
states(DOS) becomes larger with increasing energy as in Sr$_2$RuO$_4$, the
Cooper pairs with spins parallel to $\mathbf{H}$ are easier to form than
those with anti-parallel spins. This effect is especially strong in Sr$_2$RuO$_4$ due to the
Van Hove singularity just above the Fermi energy. Therefore, the
spin-polarized Cooper pairs, i.e., $k_y(\mathbf{\hat{z}}-i\alpha\mathbf{\hat{y}})$
state becomes stable near $H_{c2}$, at the expense of part of spin-orbit
coupling energy. As the magnetic field is lowered, however, the
$k_y(\mathbf{\hat{z}}-i\alpha\mathbf{\hat{y}})$ state becomes unstable,
since a non-unitary state generally loses condensation
energy. As a result, a crossover to the unitary $k_y\mathbf{\hat{z}}$ state occurs
accompanying a large entropy release. This crossover naturally leads to
the double peaks of specific heat as observed. 

We also explain the disappearance of double peaks near $T = T_c$ at low
magnetic fields in the same mechanism. The crossover is observable only when the energy gain due to
spin-polarization of Cooper pairs surpasses the spin-orbit
coupling energy. While the spin-polarization becomes stronger with
increasing magnetic field, the spin-orbit coupling energy does not change with
field. Therefore, the double peaks are observed only at low temperatures
when $H_{c2}$ is high enough.

In the following, we demonstrate our ideas given above by using the following
effective Hamiltonian,

\begin{eqnarray}
H = \sum\limits_{\mathbf{k},s}(\xi_{\mathbf{k}} - \frac{1}{2}g\mu_B\mathbf{s}\cdot\mathbf{H})C_{\mathbf{k},s}^{\dagger}C_{\mathbf{k},s} \hspace{7cm} \nonumber \\
- \frac{1}{2}g_1\sum\limits_{\mathbf{k},\mathbf{k'},s}\Bigl[\bigl(\Phi_x(\mathbf{k})\Phi_x(\mathbf{k'}) + 
\Phi_y(\mathbf{k})\Phi_y(\mathbf{k'})\bigr)C_{\mathbf{k},s}^{\dagger}C_{\mathbf{-k},s}^{\dagger}C_{\mathbf{-k'},s}C_{\mathbf{k'},s}\Bigr] \hspace{3cm} \nonumber \\
- \frac{1}{2}g_2\sum\limits_{\mathbf{k},\mathbf{k'},s}\Bigl[is_z\bigl(\Phi_x(\mathbf{k})\Phi_y(\mathbf{k'}) -
\Phi_y(\mathbf{k})\Phi_x(\mathbf{k'})\bigr)C_{\mathbf{k},s}^{\dagger}C_{\mathbf{-k},s}^{\dagger}C_{\mathbf{-k'},s}C_{\mathbf{k'},s}\Bigr] \hspace{3cm} \nonumber \\
- g_3\sum\limits_{\mathbf{k},\mathbf{k'}}\Bigl[\bigl(\Phi_x(\mathbf{k})\Phi_x(\mathbf{k'}) +
\Phi_y(\mathbf{k})\Phi_y(\mathbf{k'})\bigr)C_{\mathbf{k},\uparrow}^{\dagger}C_{\mathbf{-k},\downarrow}^{\dagger}C_{\mathbf{-k'},\downarrow}C_{\mathbf{k'},\uparrow}\Bigr] \hspace{3cm} \nonumber \\
- \frac{1}{2}g_4\sum\limits_{\mathbf{k},\mathbf{k'}}\Bigl[\bigl(\Phi_x(\mathbf{k})\Phi_x(\mathbf{k'}) -
\Phi_y(\mathbf{k})\Phi_y(\mathbf{k'})\bigr)C_{\mathbf{k},\uparrow}^{\dagger}C_{\mathbf{-k},\uparrow}^{\dagger}C_{\mathbf{-k'},\downarrow}C_{\mathbf{k'},\downarrow} +h.c.\Bigr]. \hspace{3cm}\label{Ham}
\end{eqnarray}
 Here $C_{\mathbf{k},s}$ represents electron annihilation operator for
 $\gamma$-band. $\mathbf{s}$ is the Pauli spin matrix and
 $\uparrow$($\downarrow$) means the spin direction (anti)parallel to
 $\hat{\mathbf{z}}$. The pairing functions  $\Phi_x(\mathbf{k})$ and
 $\Phi_y(\mathbf{k})$, and the coupling
 constants $g_1$-$g_4$ are obtained from the results in the third-order
 perturbation calculation on a Hubbard-like
 model\cite{rf:Yanase03}. $\Phi_x(\mathbf{k})$ and $\Phi_y(\mathbf{k})$
 have the same rotational properties as k$_x$ and k$_y$ under the
 $D_{4h}$ point group of Sr$_2$RuO$_4$. The detailed forms
 of these function do not matter in the subsequent analysis. 
The kinetic energy term $\xi_{\mathbf{k}}$ is expressed with the two-dimensional
tight-binding model for the $\gamma$-band, $\xi_{\mathbf{k}} = -2t_z(\cos(k_x) + \cos(k_y))
 - 4t_z'\cos(k_x)\cos(k_y) - \mu_z$, with $\frac{t_z'}{t_z}=0.4$ and
 $\mu_z \sim 1.555$. Note that we are interested in a
 sufficiently high magnetic field where contributions from the $\alpha$- and
 $\beta$- bands are negligible. 

Values of $g_j$ are determined in order to reproduce the transition
temperature, $T_c$, of each order parameter. $T_c$ of each order
parameter component is given by the combination of $g_j$ as shown in
Table.\ \ref{Tc}. When the spin-orbit coupling is absent, i.e., $\lambda = 0$,
all the states (A)-(C) are degenerate, i.e., $g_1 = g_3$ and $g_4 = 0$.
However, the spin-orbit coupling lifts this degeneracy and
(A)$(\Phi_x(\mathbf{k})\pm i\Phi_y(\mathbf{k}))\hat{\mathbf{z}}$ shows
the highest $T_c$, then $g_3$ becomes dominant. $g_3-g_1 > 0$ and $g_4$ are
 originated from the spin-orbit interaction and much smaller than
 $g_3$. $g_4$ determines the relative stability of
 (B)$\Phi_x(\mathbf{k})\hat{\mathbf{x}}\pm\Phi_y(\mathbf{k})\hat{\mathbf{y}}$ and 
(C)$\Phi_x(\mathbf{k})\hat{\mathbf{y}}\pm\Phi_y(\mathbf{k})\hat{\mathbf{x}}$.
Although the two-fold degeneracies in (B) and in (C) are lifted by a weak
interaction through the $g_2$ channel, $g_2$ will
play no role in the subsequent analysis and we put $g_2=0$ for simplicity. Below, we
use $\frac{1}{N(0)g_1} - \frac{1}{N(0)g_3}=0.001$ and
$\frac{1}{N(0)(g_1+g_4)} - \frac{1}{N(0)(g_1-g_4)}=0.0015$ as typical parameters.

\begin{table}[h]
  \begin{center}
   \begin{tabular}{|c|c|c|}\hline
    & pairing function & G \\ \hline
(A) & $(\Phi_x(\mathbf{k})\pm i\Phi_y(\mathbf{k}))\hat{\mathbf{z}}$ & g$_3$ \\ \hline
(B) & $\Phi_x(\mathbf{k})\hat{\mathbf{x}}\pm\Phi_y(\mathbf{k})\hat{\mathbf{y}}$ & g$_1-$g$_4$ \\ \hline
(C) & $\Phi_x(\mathbf{k})\hat{\mathbf{y}}\pm\Phi_y(\mathbf{k})\hat{\mathbf{x}}$ & g$_1+$g$_4$ \\ \hline
   \end{tabular}
   \caption{\label{Tc} Classification of pairing functions and the
   corresponding coupling constants. The transition temperature of each
   state is proportional to $\exp\bigl[-\frac{1}{N(0)G}\bigr]$ with
   N($\epsilon$) being the density of states at $\epsilon = \epsilon_F$.
}
  \end{center}
\end{table}

\begin{figure}
\includegraphics[width=0.3\textwidth]{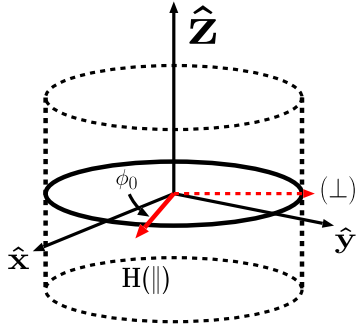}
\caption{\label{direction} The direction of magnetic field $\mathbf{H}$. The
 direction parallel to $\mathbf{H}$ is termed as $\parallel$. The
 direction perpendicular to both $\mathbf{H}$ and $\hat{\mathbf{z}}$ is termed as $\perp$.
}
\end{figure}

In order to study the superconducting state close to $H_{c2}$, we derive
Ginzburg-Landau free energy from eq.\ (\ref{Ham}) in the
weak-coupling approximation, as is legitimate for Sr$_2$RuO$_4$
($T_c/\epsilon_F\sim 10^{-4}$). We
ignore the {\bf d}-vector component parallel to
$\mathbf{H}=(H\cos\phi_0, H\sin\phi_0, 0)$, which is suppressed due to
the Pauli-paramagnetic effect. Furthermore, we only consider the
components with short coherence lengths, i.e., those with the pairing
function
$\Phi_{\parallel}(\mathbf{k})\equiv\cos\phi_0\Phi_y(\mathbf{k})+\sin\phi_0\Phi_x(\mathbf{k})$\cite{rf:Comment01}.
By taking the spin axis parallel to $\mathbf{H}$, we obtain

\begin{eqnarray}
\frac{F_{GL}}{N(0)}=t(\mid d_{\perp}\mid^2+\mid d_z\mid^2) + \alpha_{so}\mid d_{\perp}\mid^2 
+ C_{2\perp}\bigl[\mid\partial_{\perp} d_{\perp}\mid^2 + \mid\partial_{\perp} d_z\mid^2\bigr] + C_{2z}\bigl[\mid\partial_z d_{\perp}\mid^2 + \mid\partial_z d_z\mid^2\bigr] \nonumber \\ 
- i\alpha_{sp}(d_{\perp}d_z^*-d_zd_{\perp}^*) + \frac{C_4}{2}\Bigl[(\mid d_{\perp}\mid^2+\mid d_z\mid^2)^2 + \mid(d_{\perp}d_z^*-d_zd_{\perp}^*)\mid^2\Bigr], \label{gl}
\end{eqnarray}
\begin{eqnarray}
\alpha_{so}=\frac{1}{N(0)}\bigl(\frac{1}{g_1+g_4\cos^2(2\phi_0)} - \frac{1}{g_3}\bigr), \label{so}
\end{eqnarray}
\begin{eqnarray}
\alpha_{sp}=\frac{1}{4N(0)}\int\limits d\xi_{\mathbf{k}} \bigl[N(\xi_{\mathbf{k}}+\frac{1}{2}g\mu_BH)-N(\xi_{\mathbf{k}}-\frac{1}{2}g\mu_BH)\bigr] \frac{\tanh(\frac{\xi_{\mathbf{k}}}{2T})}{\xi_{\mathbf{k}}} \hspace{1cm}(>0),
\end{eqnarray}
where the order parameter $d_{\perp}$ and $d_z$ are defined as
\begin{eqnarray}
d_{\perp}=\frac{1}{2i}(\Delta_{\uparrow\uparrow}+\Delta_{\downarrow\downarrow}),
\end{eqnarray}
\begin{eqnarray}
d_z=\frac{1}{2}(\Delta_{\uparrow\uparrow}-\Delta_{\downarrow\downarrow}).
\end{eqnarray}
Here, $\Delta_{\uparrow\uparrow(\downarrow\downarrow)}$ means the order
parameter component for the electron pairs with (anti)parallel spins to the magnetic field
$\mathbf{H}$. The gauge invariant differential operator is given by
$\partial_j = \nabla_j - i\frac{2e}{c}A_j$ ($j=\perp, z$), where the
subscript $\perp$ means the direction perpendicular to both $\mathbf{H}$
and $\mathbf{\hat{z}}$ (see Fig.\ \ref{direction}). The coefficients in
eq. (\ref{gl}) are $t = \frac{T-T_{c0}}{T_{c0}}$ with $T_{c0}$ being the
transition temperature at $\mathbf{H} = 0$ for (A), and $C_{2z(\perp)} =
\frac{7\zeta(3)}{32\pi^2T_{c0}^2}\langle
v_{Fz(\perp)}^2\Phi_{\parallel}(\mathbf{k})^2 \rangle_{FS}$, $C_4 =
\frac{7\zeta(3)}{16\pi^2T_{c0}^2}\langle\Phi_{\parallel}(\mathbf{k})^4
\rangle_{FS}$ with $\langle\cdots\rangle_{FS}$ being the average over
the Fermi surface. 

The two parameters $\alpha_{so}$ and $\alpha_{sp}$ play
important roles in the subsequent analysis. $\alpha_{so}$ is originated
from the spin-orbit interaction, which stabilizes $d_z$ compared with
$d_{\perp}$. On the other hand, $\alpha_{sp}$ stems from the
spin-polarizing effect, which favors $\Delta_{\uparrow\uparrow}$
compared with $\Delta_{\downarrow\downarrow}$. To understand this, it is
helpful to rewrite this term as
$-i\alpha_{sp}(d_{\perp}d_z^*-d_zd_{\perp}^*) =
-\frac{1}{2}\alpha_{sp}(\mid\Delta_{\uparrow\uparrow}\mid^2-\mid\Delta_{\downarrow\downarrow}\mid^2)$.
As explained in the Introduction, this term comes from the
spin-dependent chemical potential shift due to the gradient of density
of states. For the spatial dependence of the order parameters, we
introduce the parameters $r$, $\theta$, and $\chi$, and put d$_{z}=r\sin\theta\Psi(\mathbf{r})$ and d$_{\perp}=r\cos\theta e^{-i\chi}\Psi(\mathbf{r})$ ($0\leq\theta\leq\frac{\pi}{2}$), where $\Psi(\mathbf{r})$ is the wave
function in the lowest Landau level, satisfying the relation
$(C_{2\perp}\partial_{\perp}^2 +
C_{2z}\partial_{z}^2)\Psi(\mathbf{r})=\sqrt{C_{2\perp}C_{2z}}\frac{2e}{c}H\Psi(\mathbf{r})\equiv
\frac{H}{H_0}\Psi(\mathbf{r})$ and being normalized as
$\int\limits\frac{d\mathbf{r}}{V}\mid\Psi(\mathbf{r})\mid^2=1$. Here, $H_0$ is
the upper-critical field at $T =0$ when $\alpha_{sp}=0$. Considering that
$H_0\sim 1.5$ Tesla for Sr$_2$RuO$_4$, we set $\frac{g\mu_BH_0}{2T_{c0}}=0.5$.
Using these order parameters, we obtain the spatial average of $F_{GL}$
($\bar{F}_{GL}$($r$,$\theta$,$\chi$)) and minimize
$\bar{F}_{GL}$ with respect to $r$ and $\chi$. The results are

\begin{eqnarray}
\bar{F}_{GL}(r_{min},\theta,\chi_{min})\propto -\frac{(t_c - t - \sqrt{(2\alpha_{sp})^2+\alpha_{so}^2}\sin^2(\theta-\theta_c))^2}{1+\sin^2(2\theta)}, \label{freeen}
\end{eqnarray}
\begin{eqnarray}
r_{min}^2 = \frac{\mid t\mid - \alpha_{so}\cos^2\theta + \alpha_{sp}\sin(2\theta) - \frac{H}{H_0}}{C_4\mid\bar{\Psi}\mid^4(1+\sin^2(2\theta))}, \label{r}
\end{eqnarray}
\begin{eqnarray}
\chi_{min} = \frac{\pi}{2},
\end{eqnarray}
where
\begin{eqnarray}
\theta_c = \frac{\pi}{4} + \frac{1}{2}\tan^{-1}\bigl(\frac{\alpha_{so}}{2\alpha_{sp}}\bigr), \label{theta_c}
\end{eqnarray}
\begin{eqnarray}
t_c = -\frac{H}{H_0} + \sqrt{\alpha_{sp}^2 + (\frac{\alpha_{so}}{2})^2} - \frac{\alpha_{so}}{2}. \label{t_c}
\end{eqnarray}
The critical temperature $t=t_c$ was found by putting $r^2=0$ and
 $\theta=\theta_c$ in eq.\ (\ref{r}). Here, $\theta_c$ expresses the spin-polarization of
Cooper pairs at $t=t_c$, and its value is determined from the competition
between $\alpha_{so}$ and $\alpha_{sp}$ as in eq.\ (\ref{theta_c}). If
spin-orbit coupling is strong enough
($\alpha_{so}\gg\alpha_{sp}$), $\theta_c$ becomes
$\sim\frac{\pi}{2}$, i.e., the state approaches a unitary state
$\Phi_{\parallel}(\mathbf{k})\hat{\mathbf{z}}$. Whereas,
if spin-orbit coupling is sufficiently weak
($\alpha_{so}\leq\alpha_{sp}$), $\theta_c$ deviates from
$\frac{\pi}{2}$ and then a non-unitary state
$\Phi_{\parallel}(\mathbf{k})(\hat{\mathbf{z}}+i\alpha\hat{\mathbf{y}}$)
 is formed. Below $t_c$, free energy is obtained by
minimizing eq.\ (\ref{freeen}) with respect to $\theta$. From eq.\
(\ref{freeen}), one can see that $\theta$ approaches $\frac{\pi}{2}$
 much below $t_c$, and the unitary state
 $\Phi_{\parallel}(\mathbf{k})\hat{\mathbf{z}}$ is formed irrespective
 of the value of $\alpha_{sp}$.

\begin{figure}
\includegraphics[width=0.45\textwidth]{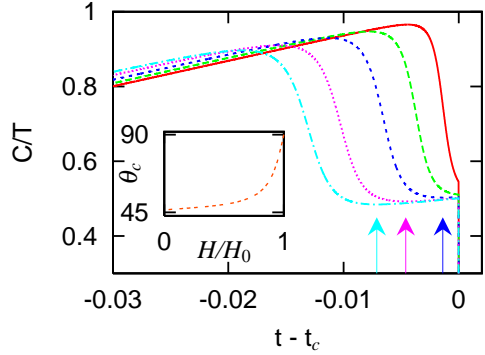} 
\caption{\label{C} Temperature dependence of the specific heat for
 $\frac{H}{H_0}=$0.20, 0.40, 0.60, 0.80, 0.90 from right to
 left. $\frac{C}{T}$ is normalized so that the jump at $t=t_c$ is 1 when
 $\alpha_{sp}=0$. The crossover temperature, $t^*$, are shown with arrows for $\frac{H}{H_0}=$0.60,
 0.80, and 0.90. For other fields, $t^*$ is absent. The inset shows the $H$
 dependence of $\theta_c$.
}
\end{figure}

\begin{figure}
\includegraphics[width=0.45\textwidth]{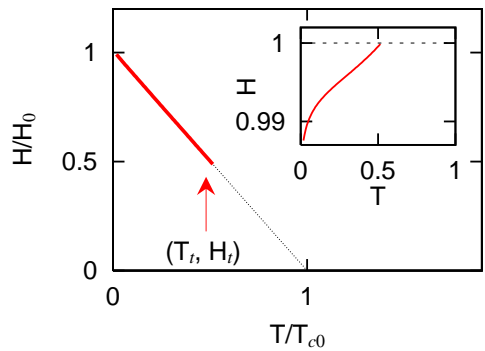}
\caption{\label{CH} H$^*$($T$) line is shown with a red line in the $T$-$H$
 diagram. A black line shows $H_{c2}$ line. The ``tricritical point'',
 ($T_t$, $H_t$), is shown with an arrow. (Inset) Same with the main
 panel, with $H$ normalized with $H_{c2}$ of each temperature. 
}
\end{figure}

From the minimized free energy, we obtain the temperature dependence of the specific
heat $C$ for various fields, which is shown in Fig.\ \ref{C}. Here, in
order to obtain $C$ correctly to the order of
$t_c - t$, we have perturbatively included the sixth order term
$\frac{31\zeta(5)\langle\Phi_{\parallel}(\mathbf{k})^6\rangle}{32\pi^4T_{c0}^4}\mid\bar{\Psi}\mid^6r^6(1+3\sin^22\theta)$
in the free energy. As shown in Fig.\ \ref{C}, $C$ has two peaks at high
magnetic fields, while only a single peak at low fields, consistent with the experiments. 
In order to characterize the crossover region, we define a crossover
temperature $t^*$ and a crossover magnetic field $H^*$ where $C$ takes
the minimum. $t^*$s are shown with arrows in Fig.\ \ref{C}. $t^*$($H^*$) can be associated
with the starting point of the crossover with decreasing temperature
(field). In Fig.\ \ref{CH}, we plot $H^*$($T$)
in $T$-$H$ diagram. $H^*$($T$) line is located just below $H_{c2}$, and is
terminated at $T\sim T_t=0.50T_{c0}$. Since the double peak structure
becomes unobservable below $T_t$, this point can be naturally
identified with the observed ``tricritical point'' as shown by an arrow
in Fig.\ \ref{CH}. This point corresponds to ($T_t$, $H_t^*$)$=$(1.2K,
0.8T).

As to the field-angle dependence of the crossover behavior, we could not
find any marked differences between $\phi_0=0$ and
$\phi_0=\frac{\pi}{4}$. Actually, while the ac-susceptibility shows the
characteristic field-angle dependence\cite{rf:Nishizaki00,rf:Deguchi02}, the double-peak structure of
specific heat does not change with field direction, thus consistent with
our analysis.

Let us make some additional remarks. Firstly, the crossover discussed in
this paper has a lot of features in common with the A$_1-$A phase
transition in $^3$He under magnetic
field\cite{rf:Osheroff,rf:Ambegaokar,rf:Brinkman}. Due to a lack of
spin-orbit coupling, a phase transition occurs in $^3$He instead of a
crossover. In order to discuss the A$_1-$A transition of $^3$He in our context, one only
has to put $\alpha_{so}=0$ in eq.\ (\ref{gl}). Then, one can find that the
A$_1-$A transition takes place at $t=t_c-2\alpha_{sp}$. Secondly, our
result does not deny the Agterberg's transition\cite{rf:Agterberg98} at
low magnetic fields. Since the $k_y(\hat{\mathbf{z}}-i\alpha\hat{\mathbf{y}})$ state
cannot be adiabatically continued to ($k_x\pm ik_y)\hat{\mathbf{z}}$ state, there still
has to be a second-order phase transition at a finite magnetic field up
to T$=$T$_{c0}$. Actually, it was found that magnetization shows a kink
at low magnetic field, suggestive of the phase transition\cite{rf:Tenya05}.

Thirdly, we would like to comment on the zero-field state. So far, we
have assumed that the ($k_x\pm ik_y$)$\hat{\mathbf{z}}$
state appears at zero magnetic field. However, this assumption is not
necessary. For example, if the
$k_x\hat{\mathbf{x}}\pm k_y\hat{\mathbf{y}}$ is stable and is nearly
degenerate with ($k_x\pm ik_y$)$\hat{\mathbf{z}}$, one will find a
crossover from $k_y(\hat{\mathbf{y}}+i\alpha\hat{\mathbf{z}})$ to
$k_y\hat{\mathbf{y}}$, and double peaks appear in specific heat in the
same way. Finally, the second-peak line of specific heat is placed at
$H\sim 0.95H_{c2}$ at $T\sim 0$, which is lower than our theoretical
estimate $H^*\sim 0.99H_{c2}$. However, in our analysis, we have not considered the
renormalization of the effective mass($m^*$) and g-factor($g^*$). Due to the electron
correlation, these factors are enhanced compared with the bare value as
$\frac{m^*}{m_{band}}\sim 5.5$ and $\frac{g^*}{g_{bare}}\sim 1.5$, respectively. Since
$\alpha_{sp}$ is proportional to the product $m^*g^*$, the
renormalization of these values will push the crossover line down to
lower field. Hence, the observed second-peak line can be reproduced even
quantitatively.

When a magnetic field is applied slightly off the plane, it is
observed that the double peaks of the specific heat disappear\cite{rf:Deguchi02}.
Currently, our theory cannot explain this phenomenon. However, the
deviation of H$_{c2}$ line from the result of Ginzburg-Landau
anisotropic effective mass model implies that we need to take account of the
layered structure of this material in order to reproduce this phenomenon. We
would like to leave this problem for future task.

In summary, we have shown that the observed double peaks of specific
heat in Sr$_2$RuO$_4$ can be attributed to the crossover from
$k_y(\hat{\mathbf{z}}-i\alpha\hat{\mathbf{y}})$ to $k_y\hat{\mathbf{z}}$
state. Within our theory, we could also explain the disappearance of
double peaks at low magnetic fields. 

We would like to thank M.\ Ichioka, M.\ Arai, and S.\ Yoshio for
valuable discussions.

\begin{thebibliography}{9}
\bibitem{rf:Comment01} Since the Fermi surface is not round, the pairing
	function with the shortest coherence length is not
	$\Psi_{\parallel}(\mathbf{k})$. However, we ignore the error
	arising from this.
\end{thebibliography}
\end{document}